\documentclass[aps,prd,twocolumn,nofootinbib,superscriptaddress,preprintnumbers,longbibliography]{revtex4-2}
\usepackage[normalem]{ulem}
\usepackage[utf8]{inputenc} 
\usepackage[T1]{fontenc}
\usepackage{ucs}
\usepackage{epsfig}
\usepackage{graphicx}
\usepackage[english]{babel}
\usepackage{hyphenat}
\usepackage{amsmath}
\usepackage{amssymb}
\usepackage{bbold}
\usepackage{mathtools}
\usepackage{mathrsfs}
\usepackage{slashed}
\usepackage{epstopdf}
\usepackage[dvipsnames]{xcolor}
\usepackage{braket}
\usepackage{booktabs}
\definecolor{lcolor}{rgb}{0.,0.0,0.}
\definecolor{citcolor}{rgb}{0,0.,0.5}
\usepackage[breaklinks,colorlinks,urlcolor=blue,citecolor=blue,linkcolor=blue]{hyperref}
\usepackage{multirow}
\usepackage{ltablex}
\usepackage{soul}
\usepackage{feynmp-auto}
\usepackage{tikz}
\usetikzlibrary{decorations.pathmorphing,decorations.markings,arrows.meta,positioning,calc}

\usepackage[e]{esvect}

\usepackage{media9}
\usepackage{nicefrac}

\newcommand{\secn}[1]{Section~1}
\newcommand{\appn}[1]{Appendix~1}

\long\def\comment#1{ }

\def\and{\quad\text{and}\quad}

\def\0{{\boldsymbol 0}}
\def\1{{\boldsymbol 1}}

\def\x{{\boldsymbol x}}
\def\y{{\boldsymbol y}}

\def\0{{\boldsymbol 0}}

\renewcommand{\part}{{\rm part}}

\newcommand{\be}{\begin{equation}}
\newcommand{\ee}{\end{equation}}
\newcommand{\bes}{\begin{subequations}}
\newcommand{\ees}{\end{subequations}}
\newcommand{\bea}{\begin{eqnarray}}
\newcommand{\eea}{\end{eqnarray}}

\newcommand{\nn}{\nonumber \\}

\def\bea#1\eea{\begin{align}#1\end{align}}
\newcommand{\bef}{\begin{figure}[h!tb]\centering}
\newcommand{\eef}{\end{figure}}

\allowdisplaybreaks

\newcommand{\ketbra}[2]{\ket{#1}\bra{#2}}

\newcommand{\kket}[1]{\ket{#1}\!\rangle}

\newcommand{\kketbbra}[2]{\ket{#1}\!\rangle\langle\!\bra{#2}}

\newcommand{\trace}[1]{\mathrm{Tr}{#1}}

\begin{document}

\title{$\Lambda \bar \Lambda$ spin correlations in high-energy collisions from quantum channels:\\
an open quantum system view of hadronization}

\author{Jo\~{a}o Barata}
\email{joao.lourenco.henriques.barata@cern.ch}
\affiliation{CERN, Theoretical Physics Department, CH-1211, Geneva 23, Switzerland}

\author{Iv{\'{a}}n Cunt{\'{i}}n} 
\email{ivan.cuntin.broullon@usc.es}
\affiliation{CERN, Theoretical Physics Department, CH-1211, Geneva 23, Switzerland}
\affiliation{Instituto Galego de F{\'{i}}sica de Altas Enerx{\'{i}}as,  Universidade de Santiago de Compostela, Santiago de Compostela 15782,  Spain}

\author{Enrique Rico}
\email{enrique.rico.ortega@cern.ch}
\affiliation{CERN, Theoretical Physics Department, CH-1211, Geneva 23, Switzerland}
\affiliation{EHU Quantum Center and Department of Physical Chemistry, University of the Basque Country UPV/EHU, P.O. Box 644, 48080 Bilbao, Spain}
\affiliation{DIPC - Donostia International Physics Center, Paseo Manuel de Lardizabal 4, 20018 San Sebastián, Spain}
\affiliation{IKERBASQUE, Basque Foundation for Science, Plaza Euskadi 5, 48009 Bilbao, Spain}

\author{Bin Wu}
\email{b.wu@cern.ch}
\affiliation{Instituto Galego de F{\'{i}}sica de Altas Enerx{\'{i}}as,  Universidade de Santiago de Compostela, Santiago de Compostela 15782,  Spain}

\preprint{CERN-TH-2026-155} 

\begin{abstract}
We construct a quantum information-centered approach to describe the experimentally observed behavior of hyperon spin-pair correlations in high-energy collider experiments. The evolution of the spin density matrix of the hyperon pair is treated in the language of quantum channels, accounting both for the spin dynamics in $\mathbb{C}^2\otimes\mathbb{C}^2$ and for the pair's angular separation $\Delta R$. We show that the experimental data are consistent with an evolution under a two-qubit depolarizing channel, from which a Lindblad master equation is derived. This provides an open quantum system picture of spin dynamics during the hadronization transition, which is not naturally captured by other quantum channels, and we discuss its microscopic origins. These results show that quantum information science can offer new insights into confinement dynamics beyond the classification of entanglement in the final particle states.
\end{abstract}

\maketitle

\section{Introduction}
One of the central open challenges in Quantum Chromodynamics (QCD), the theory of the strong interaction, is the description of the confinement mechanism, responsible for binding colored partons into color-singlet hadrons. In this strongly coupled regime of QCD, analytical understanding is limited, and much of our insight has been driven by numerical lattice QCD studies~\cite{Wilson:1974sk}, see also~\cite{tHooft:1981bkw,Mandelstam:1974pi,Luscher:2002qv,Polchinski:1991ax,Aharony:2009gg}. Despite notable advances, a dynamical understanding of confinement remains incomplete. In particular, while confinement ensures that the asymptotic particle spectrum is composed of hadrons, it does not itself explain how the quantum information carried by partons is reorganized during the hadronization process.

Although it has long been hypothesized that, in high-energy scattering processes, the parton-to-hadron transition should be inherently local~\cite{Azimov:1984np,Dokshitzer:1991wu}, thereby facilitating the probabilistic modeling of hadronization in Monte Carlo event generators~\cite{Andersson:1983ia,Webber:1983if}, several examples seem to challenge this picture. One paradigmatic case is the observation of polarized hyperons produced in unpolarized proton--proton ($pp$) and proton--nucleus scattering~\cite{Bunce:1976yb,Heller:1978ty,R608:1986ltk}, which cannot be fully explained from perturbative treatments alone~\cite{Kane:1978nd}. As such, it has since been argued that this phenomenon must involve non-perturbative spin-dependent dynamics during the hadronization stage~\cite{Anselmino:2000vs,Andersson:1979ui,DeGrand:1981du,Szwed:1981vq}. Understanding how such spin correlations are dynamically generated, therefore, provides new insight into the hadronization transition, in direct connection with the generation and survival of quantum correlations in final-state hadrons~\cite{Ellis:2011cr,Gong:2021bcp,Barata:2023qds,Liu:2026ees,Wang:2026mgc}. Nevertheless, questions involving real-time dynamics are beyond the direct scope of lattice QCD, whereas perturbative methods can address them only partially. As a result, novel analysis tools are necessary to characterize these phenomena.

In this paper, we present a novel analysis of $\Lambda$-hyperon spin correlations in $pp$ collisions from an quantum information science perspective. Under the simplifying assumption that most of the hyperon spin is carried by the heavier strange quark, we analyze the evolution of spin correlations between $\Lambda\bar{\Lambda}$ pairs in the framework of quantum channels. This allows for an agnostic treatment of the time evolution of the spin density matrix, in which the unobserved degrees of freedom involved in the hadronization transition are treated as an effective environmental bath. The resulting description naturally leads to an open quantum system picture for the loss of spin information during hadronization, accounting for its spacetime dependence. This analysis is motivated by the recent measurement of $\Lambda\bar{\Lambda}$ spin correlations in $pp$ collisions by the STAR and CMS collaborations~\cite{STAR:2025njp,CMS2026}. It was observed that, at smaller energies at STAR, spin correlations are enhanced for small angular separations, while they are suppressed at larger angles; the angular evolution at CMS energies is flatter. This behavior suggests that partonic spin information can survive the hadronization transition over short distances but is progressively degraded as the two hyperons probe less correlated regions of the QCD environment.

The main goal of this manuscript is to show that the experimentally observed behavior can be explained from a quantum information perspective. Our treatment extends the current literature on the topic in two ways. First, we incorporate the dependence of the spin-pair correlations on their spacetime separation, going beyond existing analysis of the spin structure alone~\cite{Gong:2021bcp,vonKuk:2025kbv,Liu:2026ees}. Second, we extend the more common analysis on the entanglement structure of the final state by instead emphasizing the quantum map connecting the initial partonic spin state to the observed hyperon spin density matrix. The resulting framework provides a quantitative language for describing how confinement dynamics modify partonic quantum information before they convert into final-state hadrons.

The paper is organized as follows. In Section~\ref{sec:channels}, we review aspects of the quantum channel and the open quantum system frameworks used throughout the work. In Section~\ref{sec:hyperons}, we discuss $\Lambda\bar{\Lambda}$ spin correlations in high-energy collisions, and follow to construct a microscopic model for their evolution in terms of the experimentally measured angular-dependent suppression of their spin correlations. We conclude in Section~\ref{sec:conclusion} with a discussion of the implications for quantum information based probes of hadronization and confinement. Additional discussion on the quantum channel treatment of spin correlations is provided in the Appendix.

\section{Quantum channels and open quantum systems}\label{sec:channels}
In this section, we review some basic properties of quantum channels and their connection to open quantum systems. The discussion will be restricted to the points of interest in the application in the study of $\Lambda\bar \Lambda$ spin correlations, and we refer an interested reader to~\cite{NielsenChuang, Preskill:2018,gardiner2004quantum} for a more detailed presentation of these topics.

\subsection{Quantum channels}
The state of a quantum system is described by a density operator $\rho$, which is a positive semidefinite operator of unit trace acting in the Hilbert space of the system. Although the evolution of an isolated system is unitary, many physical processes of interest involve measurements, interactions with unobserved degrees of freedom, or the loss of information to an external environment. Such processes cannot, in general, be described by unitary transformations acting solely on the system. 

The general, i.e., unitary or non-unitary, evolution of a quantum state can be understood within the formalism of quantum channels. In this context, one considers that any physical transformation of $\rho$ can be described by a completely positive map, that is, a map $\mathcal{E}: \rho \to \rho'$, such that $\mathcal{E}\otimes\mathbb{1}$ is positive in all states acting on an extended Hilbert space $\mathcal{H}\otimes\mathcal{H}'$. A quantum channel is therefore a completely positive and trace-preserving linear map that transforms an initial density matrix $\rho$ into a final state $\mathcal{E}(\rho)$. A fundamental result of quantum information theory is that every physically allowed quantum operation admits an operator-sum representation,
\begin{align}
\mathcal{E}(\rho)=
\sum_a K_a \rho K_a^\dagger\, ,
\end{align}
where the operators $\{K_a\}$ are known as Kraus operators or operation elements. Unitary evolution corresponds to the special case where there is only one term in the operator sum. Trace preservation requires
\begin{align}\label{eq:complete}
\sum_a K_a^\dagger K_a=\mathbb{1}\, ,
\end{align}
while complete positivity guarantees that the map remains positive even when acting on part of a larger entangled system. 

One can associate different choices of Kraus operators with distinct physical processes. In the context of quantum information, several choices are motivated by typical decoherence processes that occur in quantum devices, such as dephasing, amplitude damping, or depolarization~\cite{NielsenChuang, Preskill:2018,gardiner2004quantum}, although they are not unique. For example, for a single-qubit system, the depolarizing channel can be defined by
\begin{align} \label{eq:SingleQuibitDC}
    \mathcal{E}(\rho)=
    (1-p)\rho + \frac{p}{3} \sum_{i=1}^3{\sigma_i \rho \sigma_i},
\end{align}
where $\sigma_i$ denote the Pauli matrices and the parameter $p$, satisfying $0\leq p\leq 1$, quantifies the strength of the decoherence process. For $p=0$, the state remains unchanged, and the channel is the identity channel. Increasing values of $p$ gradually degrade the information encoded in the qubit. In the limit $p=\frac34$, the state is transformed into a completely mixed state. One possible Kraus representation of the depolarizing channel is given by
\begin{align}
    K_0 = \sqrt{1-p}\mathbb{1} \quad,\quad K_i = \sqrt{\frac{p}{3}}\sigma_i.
\end{align}
These operators satisfy the completeness condition, i.e., Eq.~\eqref{eq:complete}. Physically, this channel can be understood as a process in which a qubit undergoes, with equal probability, one of the three Pauli errors. The resulting dynamics are isotropic in spin space and therefore provide a simple model for the degradation of spin correlations. 

\subsection{Microscopic derivation of open quantum system master evolution equations} 
\label{sec:microscopic}

The operator-sum representation provides a general description of the evolution of open quantum systems. However, in many physical situations, it is desirable to derive the dynamics from a microscopic model describing the interaction between the system and its environment. This derivation establishes a direct connection between the properties of the environment and the effective quantum channels governing the reduced dynamics of the system.

To that end, let us consider a quantum system $S$ coupled to an environment $B$. The total Hamiltonian of $S+B$ can be written as~\cite{Breuer:2002pc}
\begin{align} \label{eq:TotalHamiltonian}
H = H_S + H_B + H_I \, ,
\end{align}
where $H_S$ and $H_B$ denote the free Hamiltonians of the system and bath, respectively, while $H_I$ describes their interaction. The state of the combined system is represented by the density operator $\rho_{\mathrm{tot}}(t)$, which evolves according to the von Neumann equation in the interaction picture
\begin{align}
\frac{d\rho_{\mathrm{tot}}(t)}{dt}=-i\left[H_I(t),\rho_{\mathrm{tot}}(t)\right].
\end{align}
Writing the interaction Hamiltonian as
\begin{align}
H_I=
\sum_\alpha A_\alpha\otimes B_\alpha\, , 
\end{align}
where $A_\alpha$ and $B_\alpha$ are system and bath operators, respectively, we decompose the system operators into eigenoperators of the free system Hamiltonian. Introducing the spectral decomposition
\begin{align}
A_\alpha
=
\sum_\omega
A_\alpha(\omega)\, ,
\end{align}
the operators $A_\alpha(\omega)$ satisfy
\begin{align}
[H_S,A_\alpha(\omega)]
=
-\omega
A_\alpha(\omega) \,,
\end{align}
where $\omega=\epsilon-\epsilon'$ is a fixed energy difference of the system, with $\epsilon$ and $\epsilon'$ being eigenvalues of $H_S$. It follows that
\begin{align}
H_I(t)=
\sum_{\alpha,\omega} e^{-i\omega t} A_\alpha(\omega)\otimes B_\alpha(t)\, .
\end{align}
The evolution equation for the reduced density matrix of the system is obtained by tracing over the environmental degrees of freedom, $\rho(t)=\mathrm{Tr}_B\left[\rho_{\mathrm{tot}}(t)\right]$. Further progress requires several physically motivated approximations.

The first of these is the \emph{Born approximation}, which assumes that the coupling between the system and the environment is sufficiently weak so that the state of the bath remains essentially unaffected by the interaction. Under this assumption, the total density matrix can be approximated at all times by a factorized form,
\begin{equation}
\rho_{\mathrm{tot}}(t)\approx \rho(t)\otimes\rho_B,
\end{equation}
where $\rho_B$ is a stationary state of the environment satisfying $[H_B,\rho_B]=0$. Physically, this means that the bath acts as a large reservoir that continuously influences the system while remaining unchanged.

The second key assumption is the \emph{Markov approximation}. The effect of the environment on the system is determined by the bath correlation functions,
\begin{equation}
\langle B_\alpha(t)B_\beta(t-s)\rangle,
\end{equation}
which quantify how long the bath retains memory of its previous state. If these correlations decay on a timescale much shorter than the characteristic evolution time of the system, the environment effectively loses memory of past interactions. The future evolution of the system then depends only on its present state and not on its previous history, resulting in a time-local master equation.

After applying the Born and Markov approximations, one obtains a master equation containing terms oscillating with frequencies of the form $e^{i(\omega-\omega')t}$. These terms couple transitions associated with different frequencies of the system. When the characteristic relaxation timescale is much longer than $\tau\approx1/|\omega-\omega'|$, the rapidly oscillating contributions average to zero over the relevant timescales. Neglecting such terms constitutes the \emph{secular approximation}, which ensures that populations and coherences associated with different transition frequencies evolve independently.

The resulting Markovian Lindblad-form master equation for the microscopic model in Eq.~\eqref{eq:TotalHamiltonian}, whose detailed derivation can be found in Appendix~\ref{app:MicroscopicDerivation}, is
\begin{multline}
\label{eq:Lindblad}
    \frac{d\rho(t)}{dt} =
    -i\left[
        H_S+H_{\mathrm{LS}},\rho(t)
    \right]
    +\\
    +\sum_\omega
    \sum_{\alpha,\beta}
    \gamma_{\alpha\beta}(\omega)
    \left(
        A_\beta(\omega)\rho(t) A_\alpha^\dagger(\omega)
        -\right. \\ \left. 
        -\frac{1}{2}\left\{
            A_\alpha^\dagger(\omega)
            A_\beta(\omega),
            \rho(t)
        \right\}
    \right)
    \, ,
\end{multline}
where we returned to the Schrödinger picture. The coefficients
\begin{align} \label{eq:CorrelationFunctions}
\gamma_{\alpha\beta}(\omega)=
\int_{-\infty}^{\infty}ds\,e^{i\omega s}\braket{B_\alpha(s)B_\beta(0)}\, ,
\end{align}
are the Fourier transforms of the bath correlation functions, and set the corresponding dissipative rates. In addition to these dissipative terms, the interaction with the bath also generates a coherent correction to the system Hamiltonian. This contribution is denoted by $H_{\mathrm{LS}}$ and is commonly referred to as the Lamb-shift Hamiltonian. It describes the renormalization of the system energy levels induced by virtual interactions with the environmental degrees of freedom. In what follows, we focus on the dissipative part of the evolution, which controls the loss of spin information and determines the effective decoherence rates.

\section{Hyperon spin correlations in high energy scattering} \label{sec:hyperons}

\subsection{Accessing the $\Lambda \bar \Lambda$ spin density matrix from experiment}
The possibility of describing hadronization as a quantum channel relies on the existence of experimentally accessible observables capable of probing the spin state of the produced hadrons. In the case of $\Lambda$ hyperons, a bound state made of a strange quark and two light quarks, such observables arise naturally from their weak and parity-violating decays, which allow the spin information carried by the hyperons to be reconstructed from the angular distributions of their decay products. Thus, by reconstructing the associated spin density matrix of hyperon pairs, one can build a dynamical picture of hadronization and its relation to spin entanglement.

To this end, we first consider how the quantum numbers of the pair can be quantified at a partonic level. In particular, in what follows, we make use of the assumption that the entire spin content of the hyperons is transmitted directly from the strange quark, while the light quarks do not directly contribute to it. This assumption is motivated by the constituent quark model, in which the $u$ and $d$ quarks combine into a spin-singlet configuration and therefore do not contribute to the total spin of the hyperon. In the non-relativistic SU(6) model, the $\Lambda$ spin is therefore determined solely by the spin of the strange quark~\cite{Fayyazuddin:1994wh}. This non-relativistic picture would be achieved in the asymptotic limit where the light-to-strange quark mass ratio is vanishing, i.e., $m_{u,d}/m_s\to 0$. Of course, in reality $m_s \approx \mathcal{O}(10) (m_u+m_d)$, and thus our working assumption should be taken as a simplifying approximation which should still capture key qualitative features. We also note that in our considerations, we shall not distinguish particles from anti-particles, and thus our consideration can not separate $\Lambda \bar \Lambda$ from $\Lambda  \Lambda$ spin correlations.

Typically in QCD, states can be classified by their discrete quantum numbers; for example, the QCD vacuum has quantum numbers $J^{PC} = 0^{++}$, where $J$ is the total angular momentum, $P$ is the parity, and $C$ is the charge conjugation. These quantum numbers are related by the following standard relations
\begin{align}
        &P=(-1)^{L+1} \, , \nn
        &C=(-1)^{L+S} \, , \nn
        &J=|L-S|, |L-S| +1, \dots, L+S \, .
\end{align}
If one assumes that the initial strange quark pair carries the quantum numbers of the vacuum, as advocated in~\cite{STAR:2025njp,Liu:2026ees}, one can deduce that the orbital momentum is $L=1$ and the spin is in the triplet state $S=1$. If the subsequent hadronization process preserves part of this spin information, the final-state $\Lambda\bar{\Lambda}$ pair may inherit the original quark-level correlations.

Experimentally, the spin of a $\Lambda$ hyperon is analyzed through its weak decay, $\Lambda\rightarrow p\pi^-$. Because the weak interaction violates parity, the angular distribution of the decay products is anisotropic and depends on the polarization of the decaying particle. For an individual hyperon, the decay distribution in its rest frame is given by
\begin{align} \label{eq:LambdaDecayDistribution}
\frac{dN}{d\Omega_p}=
\frac{1}{4\pi}\left(
    1+\alpha_\Lambda\mathbf{P}_\Lambda\cdot{\mathbf{n}_p}
\right),
\end{align}
where $\mathbf{n}_p$ is the direction of the decay proton in the rest frame of the parent particle, $\mathbf{P}_\Lambda$ is the polarization vector of the hyperon, and $\alpha_\Lambda$ is the weak decay parameter. As individual hyperons, after averaging over events, are not in a polarized state (for unpolarized proton scattering), one expects $\mathbf{P}_\Lambda=0$. However, for a $\Lambda\bar\Lambda$ system, the most general spin-density matrix compatible with the symmetries of QCD is
\begin{align} \label{eq:FinalRho}
    \rho_{\Lambda\bar\Lambda}  =
    \frac{1}{4} \left( 
        \mathbb{1} + \sum_{i=1}^3{C_{i} \sigma_i \otimes \sigma_i }
    \right), 
\end{align}
with $C_x=C_y=D_{TT}$ and $C_z=D_{LL}$, where the longitudinal direction is defined with respect to the direction of propagation in the $\Lambda\bar\Lambda$ pair center-of-mass frame, and we already imposed that each hyperon is unpolarized on average. Spin correlations are captured by the correlation matrix $C_i$, which can be extracted from the joint distribution of their decay products
\begin{align} \label{eq:JointDistribution}
\frac{dN}{d\Omega_pd\Omega_{\bar p}}=
\frac{1}{(4\pi)^2}\left(
    1+\alpha_\Lambda\alpha_{\bar\Lambda}D_{LL}\cos{\theta_p}\cos{\theta_{\bar p}} +
    \right. \\\notag \left. 
    +\alpha_\Lambda\alpha_{\bar\Lambda}D_{TT}\sin{\theta_p}\sin{\theta_{\bar p}}\cos{(\phi_p-\phi_{\bar p})}
\right),
\end{align}
where $\theta_p$ and $\phi_p$ are the polar coordinates of $\mathbf{n}_p$ in the $\Lambda$ rest frame, and $\theta_{\bar p}$ and $\phi_{\bar p}$ are defined
analogously in the $\bar{\Lambda}$ rest frame. So far, only the azimuthal average cross-section ($\cos{\theta^*} = \mathbf{n}_p\cdot\mathbf{n}_{\bar p }$)
\begin{align}
\frac{1}{N}\frac{dN}{d\cos{\theta^*}} = 
\frac12 ( 1 + \alpha_{\Lambda}\alpha_{\bar\Lambda}P_{\Lambda\Lambda}\cos{\theta^*} )
\end{align}
has been measured~\cite{STAR:2025njp}, and thus the full density matrix can not be fully fixed. Only the trace of the correlation matrix can be determined
\begin{align} \label{eq:PLambda}
P_{\Lambda\bar{\Lambda}}=\frac{1}{3}\trace{(C)}=\frac{2D_{TT} + D_{LL}}{3}.
\end{align}
Thus, in general, the final density matrix is characterized by two undetermined quantities: the initial state density matrix (which can be in a simple product state or a highly entangled spin state), and the trace of the correlation matrix of the final state.

The measurement of $P_{\Lambda\bar{\Lambda}}$ constitutes the central experimental result underlying the present work. The STAR~\cite{STAR:2025njp} and CMS~\cite{CMS2026} collaborations observed that this quantity exhibits a dependence on the relative separation of the two hyperons, giving a spatial modulation to the shape of the spin correlations, at center of mass energies $200\,\mathrm{GeV}$ and $13\,\mathrm{TeV}$, respectively. The STAR result is rather remarkable, as it suggests that for pairs produced at smaller angular distances $\Delta R=\sqrt{(\Delta y)^2+(\Delta\phi)^2}$, the extracted correlations remain sizable, indicating that a significant fraction of the initial spin information, as predicted by the SU(6) model, survives hadronization. In contrast, for larger values of $\Delta R$, the measured correlations decrease and become consistent with zero within experimental uncertainties. The CMS results are consistent with $P_{\Lambda \bar \Lambda}\approx 0$, indicating that the measured hadrons are always uncorrelated. Such observations are not inconsistent, as the spin correlations are modified during the hadronization epoch, where the interaction between the strange quarks and the bath made of light degrees of freedom can indeed deeply modify the spin density matrix. As STAR and CMS operate in very distinct kinematical regions, one expects this evolution to be qualitatively very different. In what follows, we show that, in fact, such behavior can be understood within a picture where the $\Lambda$ production is understood as an open quantum system, in which the initial $s\bar{s}$ spin state evolves in the presence of a surrounding many-body environment, corresponding to the hadronization transition in QCD. As we will show, such a problem can be tackled using the formalism introduced in Section~\ref{sec:channels}.

\subsection{Open quantum system interpretation of $\Lambda \bar \Lambda$ spin correlations} \label{sec:depolarizing}
In the previous section, we discussed how the spin state of a $\Lambda\bar{\Lambda}$ pair can be described by a two-particle density matrix and how the interaction of a quantum system with its environment may lead to decoherence. We further argued that such a picture could be used to interpret the available experimental data from STAR and CMS. These experiments have observed a signal 
compatible with the survival of the quantum correlations generated at the partonic level for small angle separation between hyperons, while for larger $\Delta R$, the measured correlations are found to decrease substantially and become consistent with zero within experimental uncertainties. This behavior suggests that the transfer of spin information from the initially produced $s\bar{s}$ pair to the final-state $\Lambda\bar{\Lambda}$ system is not perfect, and that the degree of correlation retained by the hyperon pair depends on the environment experienced during hadronization.

Motivated by these observations, we model the hadronization process as a quantum channel acting on the spin state of the initially correlated strange quark--antiquark pair. In this picture, the $s\bar{s}$ spin state constitutes the subsystem of interest, while the remaining degrees of freedom involved in the formation of the final-state hadrons act as an effective environment. Interactions with this environment progressively degrade the initial quantum correlations, transforming an initially coherent state into a mixed state. Since the depolarizing channel provides one of the simplest and most general descriptions of isotropic information loss, we employ it as a phenomenological model of the hadronization-induced decoherence process. We argue below that the behavior observed in data is inconsistent with other commonly considered quantum channels, and we provide further evidence from a heavy quark effective theory perspective.

Extending the expression for the depolarizing channel in Eq.~\eqref{eq:SingleQuibitDC} to two qubits, we can write
\begin{widetext}
    \begin{align} \label{eq:two-qubit-channel}
    \mathcal{E}(\rho) &= \sum_{\mu,\nu=0}^3{(K_{\mu}\otimes K'_{\nu})\rho (K_{\mu}^{\dagger}\otimes K_{\nu}^{\prime\dagger})} =
    (1-p)(1-p')\rho + 
    \frac{(1-p)p'}{3}\sum_{j=1}^3{(\mathbb{1}\otimes\sigma_j)\rho(\mathbb{1}\otimes\sigma_j}) \nn 
    &+
    \frac{(1-p')p}{3}\sum_{i=1}^3{(\sigma_i\otimes\mathbb{1})\rho(\sigma_i\otimes\mathbb{1})}
    +\frac{pp'}{9} \sum_{i,j=1}^3{(\sigma_i\otimes\sigma_j) \rho (\sigma_i\otimes\sigma_j})\, .
\end{align}
\end{widetext}
Applying this channel to the initial $s\bar s$ density matrix 
\begin{align} \label{eq:InitialRho}
    \rho_{s\bar s}  =
    \frac{1}{4} \left( 
        \mathbb{1} + \sum_{i}{C^{(0)}_{i} \sigma_i \otimes \sigma_i }
    \right), 
\end{align}
we directly find
\begin{align}
    \mathcal{E}(\rho_{s\bar s}) = 
     \frac{\mathbb{1}}{4}
         +\frac{1}{4}\left(1-\frac43 p\right)\left(1-\frac43 p'\right) \sum_{i}{C_i^{(0)}\sigma_i\otimes\sigma_i}.
\end{align}
That is, the channel leaves the identity invariant, damps single-qubit coherences
with factors $(1-\frac43 p)$ and $(1-\frac43 p')$, and suppresses two-qubit correlations
multiplicatively. This reflects the independent action of two depolarizing
environments on the subsystems.

While the channel formulation provides a non-differential description of the evolution, it is often useful to reformulate the same process in terms of a master equation, as will be shown in the next section.

\subsubsection{Master equation for the depolarizing channel} \label{sec:QCtoME}
As we have discussed, it is often possible to describe the (not necessarily coherent) evolution of a density operator, at least to a good approximation, by a master differential equation 
\begin{align}
    \dot\rho = \mathcal{L}(\rho) \, .
\end{align}
The first step to relate a master equation to a corresponding quantum channel $\mathcal{E}(\rho)$~\cite{Andersson:2007nas} relies on expressing both the quantum channel and the Liouvillian $\mathcal{L}$ in a matrix form:
\begin{align} \label{eq:LandFMatrices}
    &F_{\mu\kappa,\nu\lambda} = 
    \frac{1}{4}\trace{\left[
        \mathcal{G}_{\mu\kappa}\mathcal{E}(\mathcal{G}_{\nu\lambda})
    \right]} \, ,\nn
    &L_{\mu\kappa,\nu\lambda} = 
    \frac{1}{4}\trace{\left[
        \mathcal{G}_{\mu\kappa}\mathcal{L}(\mathcal{G}_{\nu\lambda})
    \right]}\,,
\end{align}
with $\mathcal{G}_{\mu\kappa}=\sigma_{\mu}\otimes\sigma_{\kappa}$ and $\sigma_0 = \mathbb{1}$. $F$ and $L$ satisfy the relation $L = \dot F F^{-1}$. As a result, this set of formulas allows us to obtain a master equation from a quantum channel and vice versa. At this point it is convenient to do the reparametrization $p\rightarrow\frac{3}{4}p$, so that the matrix representation of $\mathcal{E}$ becomes
\begin{align}
    F_{00,00} = 1 \, ,\quad F_{0i,0i} = 1 - p'\, , \quad F_{i0,i0} = 1 - p\, ,
    \\\notag
    F_{ij,ij} = pp' - (p+p') + 1 = (1-p)(1-p')\, .
\end{align}
Introducing time dependence in the probabilities as $p(t)=1-e^{-\gamma t}$ and $p'(t)=1-e^{-\gamma' t}$, with $\gamma$ and $\gamma'$ the decoherence rates, the Liouvillian matrix takes the form
\begin{align} \label{eq:DCLMatrix}
    & L_{00,00} = 0, \quad L_{0i,0i} = -\gamma', \quad L_{i0,i0} = -\gamma\, ,
\end{align}
and $L_{ij,ij} =L_{0i,0i} + L_{i0,i0}$. From $L$, one directly obtains the Liouvillian 
\begin{widetext}
   \begin{align} \label{eq:DC_Lindblad}
    \mathcal{L}(\rho) &=
    \frac{1}{4}\sum_{\mu,\kappa}\sum_{\nu,\lambda}{
        L_{\mu\kappa,\nu\lambda}\trace{(\mathcal{G}_{\mu\kappa}\rho)}\mathcal{G}_{\nu\lambda}
    }     
    =\frac{\gamma}{4}\sum_i{\left(
        \mathcal{G}_{i0}\rho\mathcal{G}_{i0}^{\dagger} - 
        \frac{1}{2}\{\mathcal{G}_{i0}\mathcal{G}_{i0}^{\dagger},\rho\}
    \right)} 
    +\frac{\gamma'}{4}\sum_i{\left(
        \mathcal{G}_{0i}\rho\mathcal{G}_{0i}^{\dagger} - 
        \frac{1}{2}\{\mathcal{G}_{0i}\mathcal{G}_{0i}^{\dagger},\rho\}
    \right)} \, .
\end{align} 
\end{widetext}
Equation~\eqref{eq:DC_Lindblad} is the Lindblad master equation corresponding to two independent single-qubit depolarizing channels acting on the two-qubit system. The dissipative dynamics are generated by the local Lindblad operators $\mathcal{G}_{i0}=\sigma_i\otimes\mathbb{1}$ and $\mathcal{G}_{0i}=\mathbb{1}\otimes\sigma_i$, which act on the quark and anti-quark subsystems, respectively.

\subsubsection{Building up the master equation from microscopics}
The phenomenological depolarizing channel introduced in the previous section provides a simple description of the degradation of spin correlations during hadronization. It would be desirable to establish a microscopic framework capable of relating the parameter of the channel, $\gamma$, to the underlying dynamics of the bath environment. In the weak-coupling and Markovian regimes, depolarizing dynamics can be derived from an explicit system--environment interaction Hamiltonian in which the spin degrees of freedom couple to bath operators describing the surrounding medium. We therefore begin by considering a microscopic model in which the spin state of the initially correlated $s\bar{s}$ pair interacts with an effective bath through operators $B_i$ acting on the environmental Hilbert space. Under the standard Born--Markov and secular approximations, this interaction gives rise to a Lindblad master equation whose solution reproduces the depolarizing channel.

While such a description accounts for the progressive loss of spin information, it does not explain the experimentally observed dependence of the spin correlations on the relative separation of the final-state hyperons. To address this limitation, we extend the model by introducing bath operators with an explicit spatial dependence. In this formulation, the environment is described by field operators evaluated at the locations associated with the two constituents of the subsystem, allowing the degree of bath correlation experienced by the pair to depend on their separation. As a consequence, the resulting decoherence rates become functions of the relative distance and, ultimately, of the experimentally accessible variable $\Delta R$.

Since the purpose of the model is to describe the evolution of the spin degrees of freedom rather than the full color dynamics of QCD, we adopt an effective bath inspired by the quantized electromagnetic field. The bath operators are therefore constructed from bosonic creation and annihilation operators in analogy with the coupling of a spin to fluctuating magnetic fields. This choice captures the essential ingredients required for spin decoherence---namely, the presence of fluctuating environmental fields and their spatial correlations---while remaining sufficiently simple to permit an analytic derivation of the corresponding master equation.

To this end, we first note that Eq.~\eqref{eq:DC_Lindblad} can be derived from an interaction Hamiltonian of the form~\cite{Fonseca_Romero_2012,Arsenijevic:2017kdx}
\begin{align} \label{eq:Hmodel}
    H_I = \sum_i{ \bigg(
        A_{i}^{(s)}\otimes B_i^{(s)} + A_{i}^{(\bar s)}\otimes B_i^{(\bar s)}\bigg)
    }\, ,
\end{align}
with $A_{i}^{(s)}= \nicefrac{\mathcal{G}_{i0}}{2}$ and $A_{i}^{(\bar s)} = \nicefrac{\mathcal{G}_{0i}}{2}$, where the bath operators are of the form
\begin{align}
    B^{(q)}_i = \int{d\omega\, g^{(q)}(\omega)(a_{i\omega} + a^{\dagger}_{i\omega}) } \, ,
\end{align}
where $g^{(q)}(\omega)$ represents the interaction strength per frequency, with $q=s,\bar s$. The system Hamiltonian is
\begin{align}
    H_S = \omega_0 ( A_{z}^{(s)} + A_z^{(\bar s)})\, .
\end{align}
Using these elements, and following the procedure outlined in Section~\ref{sec:microscopic}, one finds, for the case where $\omega_0=0$, that the evolution of the density matrix corresponds to that of Eq.~\eqref{eq:DC_Lindblad}.

We further consider an extension of this model in which the interaction Hamiltonian takes the form
\begin{align} \label{eq:Hmodel}
    H_I = \sum_i{\bigg[
        A_{i}^{(s)}\otimes B(\mathbf{x})+ A_{i}^{(\bar s)}\otimes B(\mathbf{y})
    \bigg]} \, , 
\end{align}
where the bath operator is
\begin{align}
B(\mathbf{x}) &=
\int d^3k \,\left(
    g(k) a_{\mathbf{k}} e^{i\mathbf{k}\cdot \mathbf{x}} +
    g^*(k) a_{\mathbf{k}}^\dagger e^{-i\mathbf{k}\cdot \mathbf{x}}
\right) \, ,
\end{align}
where $g(k)$ is the generally isotropic coupling strength between the system under study and the unmeasured system (taken as the bath) in momentum space, which depends only on the magnitude $k=|\mathbf{k}|$. This extension thus allows one to include spatial information in the bath and capture the angular dependence of the spin correlator. Following once again the method described in Section~\ref{sec:microscopic}, in the case where $\omega_0=0$ we find
\begin{align} \label{eq:SpatialDC}
    \frac{d\rho}{dt} &= 
    \mathcal{L}^{(\mathrm{DC})}(\rho) \nn 
    &+ 
    \frac{\xi(r,0)}{4}\sum_i{(
        \mathcal{G}_{i0}\rho\mathcal{G}_{0i} +
        \mathcal{G}_{0i}\rho\mathcal{G}_{i0} -
        \{\mathcal{G}_{ii}, \rho \})}\, .
\end{align}
Here, $\mathcal{L}^{(\mathrm{DC})}(\rho)$ is the Liouvillian for the depolarizing channel in Eq.\eqref{eq:DC_Lindblad}, with decoherence rate
\begin{align}
\gamma(\omega) &=
\int_{-\infty}^{\infty}ds\,e^{i\omega s}\braket{B(\mathbf{x},s)B(\mathbf{x},0)} 
\notag\\
&= \int_{-\infty}^{\infty}ds\,e^{i\omega s}\braket{B(\mathbf{y},s)B(\mathbf{y},0)}.
\end{align}
The second term in Eq.~(\ref{eq:SpatialDC}) accounts for correlated depolarization induced by a common environment. These non-local dissipative contributions are governed by the decoherence rate
\begin{align} 
\xi(r,\omega)=
\int_{-\infty}^{\infty}ds\,e^{i\omega s}\braket{B(\mathbf{x},s)B(\mathbf{y},0)} \, ,
\end{align}
with $r=|\mathbf{x}-\mathbf{y}|$.
Using the bath average
\begin{align}
\langle &B(\mathbf{x},s) B(\mathbf{y},0)\rangle
=
\int d^3k\, |g(k)|^2
\nn &\times\bigg[
(N(k)+1)e^{-ik s}
+
N(k)e^{ik s}
\bigg]
e^{i\mathbf{k}\cdot(\mathbf{x}-\mathbf{y})}\, ,
\end{align}
with $N(k)$ the occupation number,  we arrive at
\begin{align}
\xi(r,\omega)=\gamma(\omega)\frac{\sin(k_\omega r)}{k_\omega r}\, ,
\end{align}
with $\gamma(\omega) =  8\pi^2 k_{\omega}^2 |g(k_{\omega})|^2 N(\omega)$ and $k_{\omega}=\omega/c$. The coefficient $\gamma(0)$ corresponds to the correlation function for the same-point bath evaluated at zero frequency. In general, the low-frequency behavior of both the coupling $g(k)$ and the occupation number $N(k)$ is not known from first principles in QCD. As a result, the zero-frequency limit cannot be determined microscopically and must be treated as an effective parameter of the reduced dynamics.
\begin{align}
    \gamma(0) = 8\pi^2 \lim_{\omega\rightarrow0} k_{\omega}^2 |g(k_{\omega})|^2 N(\omega) \equiv \gamma \, ,
\end{align}
where the last equality defines the effective decoherence rate $\gamma$.
On the other hand, the separate-baths correlation function is
\begin{align}
    \xi(r) = \lim_{\omega\rightarrow0} \gamma\frac{\sin(k_\omega r)}{k_\omega r}\, .
\end{align}
Since the relevant dynamics occur in the zero-frequency limit, the quantity $k_\omega r$ is not uniquely determined by the microscopic model. Nevertheless, the correlation function must satisfy two physically motivated conditions: for vanishing separation, both spins should experience the same environmental fluctuations, implying $\xi(r)\rightarrow\gamma$ as $r\rightarrow0$; conversely, for large separations the bath correlations should disappear, leading to $\xi(r)\rightarrow0$ as $r\rightarrow\infty$. To interpolate between these two regimes, we replace $k_\omega r$ by the dimensionless variable $ar$, where $a$ determines the characteristic correlation length of the environment, which we set at $a=1$.

Thus, this microscopic model predicts correlations as a function of the spatial separation between the two hadronization points, whereas experimentally the accessible quantity is the angular separation $\Delta R$ in momentum space. Motivated by the approximately local nature of hadronization, we assume that the average spatial separation is a monotonic function of $\Delta R$, allowing the proportionality constant to be absorbed into the environmental correlation length.

\subsubsection{Numerical fit to experimental data}
Having constructed explicit forms for the bath correlation functions, parametrized by $\gamma$, we can now extract these from the experimental values of $P_{\Lambda\bar\Lambda}$. To find an analytical relation between $\xi$ and $P_{\Lambda\bar\Lambda}(\Delta R)$, we require the quantum channel corresponding to Eq.~\eqref{eq:SpatialDC}, which can be obtained following the same method already employed in Section \ref{sec:QCtoME}. We start by writing the Liouvillian in matrix form. The diagonal terms of $L$ are the same ones as in Eq.~(\ref{eq:DCLMatrix})
\begin{align}
    L_{00,00} = 0, \; L_{0i,0i} = -\gamma, \; L_{i0,i0} = -\gamma, \;
    L_{ij,ij} = -2\gamma,
\end{align}
whereas the off-diagonal terms $(i\neq j)$ can be found by using Eq.~(\ref{eq:LandFMatrices}) again and are
\begin{align}
    L_{ii,jj} = \xi(\Delta R), \quad L_{ji,ij} = - \xi(\Delta R).
\end{align}
We find that the quantum channel can be written, for an initial density matrix as in Eq.~\eqref{eq:InitialRho}, as
\begin{align}
    \rho_{\Lambda\bar{\Lambda}}&=\mathcal{E}(\rho_{s\bar s}) =
    \frac{1}{4}\sum_{\mu,\kappa}{\sum_{\nu,\lambda}{
        F_{\mu\kappa,\nu\lambda}\trace{(\mathcal{G}_{\mu\kappa}\rho_{s\bar s})}\mathcal{G}_{\nu\lambda}
    }}\nn 
    &=\frac{\mathbb{1}}{4}
        +
        \frac14\sum_{i}{
            C^{(0)}_i\left(
                \alpha
                \mathcal{G}_{ii}
                +
                \beta
                \sum_{j\neq i}{
                    \mathcal{G}_{jj}
                }
            \right)} \, ,
\end{align}
with $F=e^{Lt}$, since $L$ does not depend on time, and
\begin{align}
    \alpha = \frac13 e^{-(2\gamma-2\xi(\Delta R))t}+ \frac23 e^{-(2\gamma+\xi(\Delta R))t}\, ,
    \nn
    \beta = \frac13\left[e^{-(2\gamma-2\xi(\Delta R))t}-e^{-(2\gamma+\xi(\Delta R))t}\right]\, .
\end{align}
Connecting these to the parametrization of the measured density matrix and the initial one, we find the simple relation
\begin{align}
    D_T &= D_T^{(0)}(\alpha+\beta) + \beta D_L^{(0)} \, ,
    \nn 
    D_L &= \alpha D_L^{(0)} + 2\beta D_T^{(0)}\, .
\end{align}
As a result, the initial and final polarization parameters are connected by a simple relation:
\begin{align} \label{eq:ThPLambda}
    P_{\Lambda\bar\Lambda}(\Delta R) = 
    P^{(0)}_{\Lambda\bar\Lambda}\exp{ \left(-2\left[\gamma-\xi(\Delta R)\right]t\right) }\, .
\end{align}
Note that this evolution is derived from the structure of the quantum channel, matching the heuristic \textit{ansatz} recently used in~\cite{Liu:2026ees}.

To quantify the dependence of the spin correlations on the pair separation, we fit the measured values of $P_{\Lambda\bar{\Lambda}}$ as a function of $\Delta R$ using the phenomenological expression in Eq.~\eqref{eq:ThPLambda} where $\gamma$ denotes the local decoherence rate, $t$ is the effective interaction time, and
\begin{equation}
\xi(\Delta R)=\gamma\frac{\sin{(\Delta R)}}{\Delta R}\, ,
\end{equation}
describes the spatial correlations of the environment. For small values of $\Delta R$, the environmental fluctuations experienced by the two particles are strongly correlated, yielding $\xi(\Delta R)\simeq\gamma$ and therefore a suppression of decoherence. As the separation increases, the correlation function decreases, and the effective decoherence rate approaches its asymptotic value $\gamma$, leading to a progressive loss of spin correlations.

The value of the initial correlation $P^{(0)}_{\Lambda\bar{\Lambda}}$ depends on the assumptions used to relate the spin state of the parent $s\bar{s}$ pair to that of the observed $\Lambda\bar{\Lambda}$ system. As discussed previously, the value $P^{(0)}_{\Lambda\bar{\Lambda}}=\frac13$ follows from the non-relativistic SU(6) constituent quark model, in which the spin of the $\Lambda$ hyperon is entirely carried by the strange quark. While this picture provides a useful starting point, it is ultimately a simplifying approximation and may receive corrections from more realistic descriptions of the hadron structure. Furthermore, feed-down contributions from heavier hyperons are expected to dilute the observable correlations. For these reasons, we do not impose a fixed value of $P^{(0)}_{\Lambda\bar{\Lambda}}$ and instead treat it as a free parameter. The fit therefore contains two free parameters: the effective decoherence strength $\gamma t$ and the initial correlation $P^{(0)}_{\Lambda\bar{\Lambda}}$.

The fit is performed independently for the STAR and CMS datasets. In both cases, the first experimental point is excluded from the analysis. For the average transverse momentum of the STAR and CMS samples, the smallest measured separation corresponds to a characteristic scale
$Q \sim \langle p_T\rangle \Delta R \approx 0.34\,\mathrm{GeV}$, comparable to the non-perturbative QCD scale $\Lambda_{\mathrm{QCD}}$. In this regime, additional low-energy hadronic effects not included in the present open-system description may become relevant. Since the model is intended to provide an effective description of decoherence at scales above the non-perturbative region, the smallest-$\Delta R$ point is omitted from the fit.

The numerical fit yields the results
\begin{align}
    &\gamma t_{{\rm STAR}} =3.5\pm2.7\, , \nn &P^{(0)}_{\Lambda\bar\Lambda,\, {\rm STAR}} =0.33\pm0.21 \,, 
\end{align}
for the STAR dataset, while for the CMS data we obtain 
\begin{align}
    &\gamma t_{\rm CMS} = 7.2\pm8.6\, , \nn
    &P^{(0)}_{\Lambda\bar\Lambda,\, {\rm CMS}} = -0.11\pm0.18.
\end{align}
\begin{figure}[h!]
    \centering
    \includegraphics[width=1\linewidth]{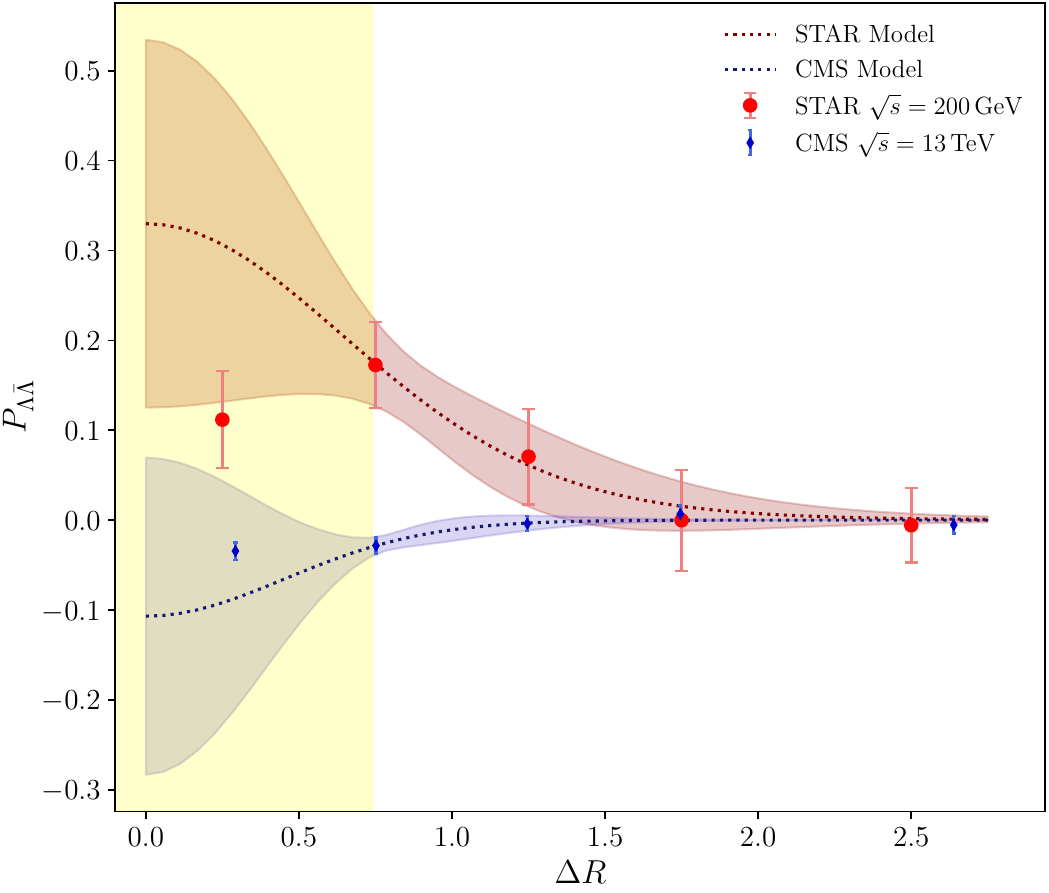}
    \caption{Spin correlations of $\Lambda\bar\Lambda$ pairs in $pp$ collisions at $\sqrt{s}=200\,\mathrm{GeV}$ (red)~\cite{STAR:2025njp} and $\sqrt{s}=13\,\mathrm{TeV}$(blue)~\cite{CMS2026}. The yellow band represents the non-perturbative regime below $\Delta R \sim \nicefrac{\Lambda_{\mathrm{QCD}}}{\braket{p_T}}$, with $\braket{p_T}^{\mathrm{CMS}}\approx\braket{p_T}^{\mathrm{STAR}}=1.35$ GeV. Dotted lines represent the fit of Eq.~\eqref{eq:ThPLambda} to the experimental values.}
    \label{fig:placeholder1}
\end{figure}
The fitted parameters exhibit large uncertainties of comparable magnitude. This reflects the limited constraining power of the dataset, which consists of a few experimental points with large uncertainties. The STAR fit provides evidence for a progressive suppression of spin correlations with increasing separation, consistent with the decoherence mechanism proposed in this work. The extracted value of $P^{(0)}_{\Lambda\bar{\Lambda}}$ is close to the prediction of the SU(6) model, although its interpretation is complicated by the expected feed-down contributions and by the exclusion of the smallest-$\Delta R$ point, whose measured value is itself close to the feed-down-diluted expectation. It is also worth noting that the overall value of $P_{\Lambda\bar{\Lambda}}=0.181\pm0.035\pm0.022$ reported by the STAR collaboration exceeds the feed-down-diluted prediction, suggesting that the relation between the observed $\Lambda\bar{\Lambda}$ correlations and the underlying $s\bar{s}$ spin state may be more intricate than assumed in the simplified picture adopted here. In contrast, the CMS data do not exhibit a statistically significant positive correlation at small separations and remain compatible with zero over the full $\Delta R$ range. As a consequence, the fitted parameters are poorly constrained and do not permit a meaningful test of the underlying production mechanism.

\subsubsection{Incompatibility with other quantum channel descriptions}
Besides the adopted depolarizing channel for the phenomenological description of hadronization-induced decoherence, several other alternative quantum channels were also considered. In particular, we studied both dephasing and amplitude-damping dynamics, which represent two of the most common mechanisms of environmental noise.

\paragraph*{Dephasing channel.} This channel suppresses quantum coherence while preserving state populations, leading to a final state that retains a preferred spin basis and therefore does not reproduce the nearly isotropic reduction of spin correlations suggested by the experimental observations. In other words, it has the effect of suppressing the off-diagonal terms while leaving the diagonal terms intact, which, for the case of a system in the initial state (see Eq.\eqref{eq:InitialRho}), amounts to the suppression of $D^{(0)}_{TT}$ while $P_{\Lambda\bar\Lambda} = \frac13(2D_{TT} + D_{LL})$ would approach $D^{(0)}_{LL}$ and not zero, as supported by experiment.

\paragraph*{Amplitude-damping channel.}This channel drives the system towards a specific ground state, introducing a directional bias and describing energy relaxation rather than the loss of spin information expected during hadronization.

Neither channel was found to generate the desired evolution from an initially correlated spin-triplet state to the approximately unpolarized final state inferred from the data. The depolarizing channel, by contrast, degrades all spin components uniformly and drives the system towards a maximally mixed state while preserving rotational symmetry. It therefore provides the simplest effective description of the observed suppression of spin correlations.

\subsubsection{Connection to heavy quark effective theory}
So far, our discussion, though motivated by a problem in QCD, has not used any element of the theory, but rather explained the observed phenomena from a purely quantum information point of view. Here, we motivate that the previous results can be connected to QCD.

To do this, we shall rely on heavy quark effective theory (HQET), which can be employed to explain the dynamics of hadron states made from heavy quarks~\cite{Isgur:1989vq,Isgur:1990yhj,Eichten:1989zv,Georgi:1990um,Grinstein:1990mj,Neubert:1996wg}. In HQET, the leading
spin-sensitive interaction between a heavy quark and the soft gluon background is the chromomagnetic Pauli operator, which is introduced in the Lagrangian via
\begin{align}
    \mathcal L_{\rm spin}
    =
    \kappa_s\,
    h_v^\dagger S^i g B_i h_v \, ,
    \label{eq:HQETSpinPauli}
\end{align}
where $h_v$ is the heavy-quark effective field with velocity $v$, 
$S^i=\sigma^i/2$ is the spin operator, $g$ is the QCD coupling, and 
$B_i$ is the chromomagnetic field. The coefficient
\(\kappa_s=c_F/m_s\) controls the strength of the spin--chromomagnetic
interaction. In the tree-level approximation, one may take \(c_F=1\), so that \(\kappa_s=1/m_s\). We note that in this set-up, although we assume the strange quarks are heavy, the previous expression should not be interpreted as a controlled
expansion in $\Lambda_{\rm QCD}/m_s$, and we use Eq.~\eqref{eq:HQETSpinPauli}
as a symmetry-motivated guide for the leading local spin-dependent coupling of the strange quark to the soft QCD environment. The purpose of the HQET
argument is therefore not to make a quantitatively controlled expansion in $1/m_s$, but to identify the chromomagnetic bath operator responsible for spin decoherence.

After the standard Wilson-line field redefinition, which removes the leading eikonal coupling $v\cdot A$ from the heavy-quark Lagrangian, the spin interaction Hamiltonian  can be written in the form
\begin{align}
    H_I(t)
    =
    -\sum_{a=1}^{2}
    S_a^i\,\mathcal O_{ai}(t),
    \label{eq:HQETSpinInteraction}
\end{align}
where $a=1,2$ labels the strange quark and antiquark trajectories, and
\begin{align}
    \mathcal O_{ai}(t)
    =
    \kappa_a\,
    Y_a^\dagger(t)\,
    g B_i^A(x_a(t))\,T_a^A\,
    Y_a(t) \, .
    \label{eq:WilsonLineChromomagneticOperator}
\end{align}
Here $Y_a$ is a Wilson line along the corresponding trajectory, while $T_a^A$ is the color generator in the representation appropriate to the quark or antiquark. The operator $\mathcal O_{ai}$ is thus the QCD analogue of the bath operator used above. Using the Born--Markov approximation and neglecting spin splittings over the short hadronization time, tracing over the soft QCD degrees of freedom gives the Lindblad equation
\begin{align}
    \frac{d\rho}{dt}
   &=
    \sum_{a,b=1}^{2}
    \Gamma_{ab}^{ij}
    \left(
        S_a^i \rho S_b^j
        -
        \frac{1}{2}
        \left\{
            S_b^j S_a^i,\rho
        \right\}
    \right)\nn 
    &-i[H_{\rm LS},\rho],
    \label{eq:HQETLindblad}
\end{align}
where 
\begin{align}
    \Gamma_{ab}^{ij}
    =
    \int_{-\infty}^{+\infty} d\tau\,
    \left\langle
        \mathcal O_{ai}(\tau)\mathcal O_{bj}(0)
    \right\rangle_{\rm bath} \, .
    \label{eq:HQETKossakowski}
\end{align}
As discussed before, the Lamb-shift Hamiltonian $H_{\rm LS}$ generates coherent spin precession; since our main interest is the decay of correlations, we focus on the dissipative part of Eq.~\eqref{eq:HQETLindblad} solely in what follows.

For an unpolarized and spin-isotropic environment, and after tracing over the unresolved color and soft degrees of freedom, the most general leading structure relevant for the scalar spin correlations can be parametrized as
\begin{align}
    \Gamma_{ab}^{ij}
    =
    \delta^{ij}\Gamma_{ab},
    \qquad
    \Gamma_{ab}
    =
    \begin{pmatrix}
        \gamma & \xi(r) \\
        \xi(r) & \gamma
    \end{pmatrix} \, ,
    \label{eq:GammaMatrixHQET}
\end{align}
with $r=|\x-\y|$. The local coefficient $\gamma$ measures the strength of the chromomagnetic spin
noise experienced by each particle separately, while the nonlocal coefficient
$\xi(r)$ measures the correlation between the noise sampled along the two
trajectories. Complete positivity requires
\begin{align}
    \gamma+\xi(r)\geq 0,
    \qquad
    \gamma-\xi(r)\geq 0,
    \qquad
    |\xi(r)|\leq \gamma .
    \label{eq:HQETPositivity}
\end{align}
It is useful to diagonalize Eq.~\eqref{eq:GammaMatrixHQET}, introducing $J_i^{(\pm)} =\frac{S_1^i\pm S_2^i}{\sqrt{2}}$, such that the system's evolution can be written as 
   \begin{align}
    \frac{d\rho}{dt}
    &=
    (\gamma+\xi)
    \sum_i
    \mathcal D[J_i^{(+)}]\rho
    \nn 
    &+
    (\gamma-\xi)
    \sum_i
    \mathcal D[J_i^{(-)}]\rho
    -i[H_{\rm LS},\rho]\,,
    \label{eq:CommonRelativeLindblad}
\end{align} 
where $\mathcal D[L]\rho = L\rho L^\dagger - \frac{1}{2} \left\{  L^\dagger L,\rho \right\}$. The common ($+$) mode corresponds to the same chromomagnetic fluctuation acting on both spins, whereas the relative ($-$) mode corresponds to fluctuations that distinguish the two trajectories. This decomposition is useful when considering the action of the adjoint Lindblad generator on the scalar spin-correlation operator, $ \boldsymbol\sigma_1\cdot\boldsymbol\sigma_2 $, for which one finds the identities 
\begin{align}
    &\sum_i
    \mathcal D^\dagger[J_i^{(+)}](\boldsymbol\sigma_1\cdot\boldsymbol\sigma_2)
    =0\, ,\nn 
    &\sum_i
    \mathcal D^\dagger[J_i^{(-)}](\boldsymbol\sigma_1\cdot\boldsymbol\sigma_2)
    =
    -2 \, \boldsymbol\sigma_1\cdot\boldsymbol\sigma_2 \, .
    \label{eq:AdjointActionSpinCorrelation}
\end{align}
Thus, only the relative chromomagnetic noise depolarizes the pair and suppresses the scalar spin correlation. One therefore obtains
\begin{align}
    P_{s\bar s}(t,r)
    =
    P_{s\bar s}^{(0)}
    \exp\left[
        -2[\gamma-\xi(r)]t
    \right].
    \label{eq:HQETSpinCorrelationSuppression}
\end{align}
This has precisely the same structure as the suppression factor obtained from
the depolarizing-channel analysis, giving a QCD interpretation of the phenomenological channel parameters. In particular, the overall decoherence strength is controlled by the local chromomagnetic spin-noise rate $\gamma$, while the angular dependence is governed by the nonlocal chromomagnetic correlator $\xi(r)$.

This simple derivation further motivates why the depolarizing channel is the natural leading channel for the present observable. An unpolarized, spin-isotropic chromomagnetic environment produces isotropic spin noise, which suppresses $\boldsymbol\sigma_1\cdot\boldsymbol\sigma_2$ without selecting a preferred spin axis. Channels such as pure dephasing or amplitude damping would require additional microscopic structure, such as a preferred quantization axis, anisotropic bath correlations, or sizable coherent spin splittings. Even though such effects can be taken into consideration, it is unclear to us how to justify their relevance in a simple physical picture.

\section{Conclusion} \label{sec:conclusion}
In this work, we have investigated the possibility that the suppression of spin correlations observed in $\Lambda\bar{\Lambda}$ production can be interpreted as a manifestation of quantum decoherence during the hadronization process. Motivated by recent measurements from the STAR and CMS collaborations, we employed the formalism of quantum channels and open quantum systems to model the evolution of the spin state of a parent $s\bar{s}$ pair into the experimentally observed $\Lambda\bar{\Lambda}$ final state.

We first considered several phenomenological quantum channels capable of describing the loss of quantum information. In particular, dephasing and amplitude-damping channels were examined and found to be incompatible with the approximately isotropic suppression of spin correlations suggested by the data. By contrast, the depolarizing channel naturally drives the system towards a maximally mixed state while preserving rotational symmetry, making it the simplest effective description of the observed behavior.

To provide a microscopic justification for this phenomenological picture, we derived a Lindblad-form master equation from a system--environment Hamiltonian. The environment was modeled as a bosonic bath coupled to the spin degrees of freedom of the strange quark and antiquark. By introducing spatially dependent bath operators, we obtained an effective decoherence model in which the decay of spin correlations depends on the separation between the two particles. This construction leads naturally to a correlation function of the form $\xi(r)$ and establishes a direct connection between the environmental correlation length and the experimentally measured variable $\Delta R$. We iterate that the inclusion of the angular dependence in the description of the spin correlations sets this approach apart from the existing literature.

The resulting model predicts an exponential suppression of the spin correlation observable $P_{\Lambda\bar{\Lambda}}$ with increasing separation. Fitting this expression to the STAR data yields a clear indication of a decreasing correlation strength as $\Delta R$ increases, consistent with the interpretation of hadronization as a decohering process. The extracted effective decoherence parameter is reasonably well constrained and reproduces the overall trend of the data. In contrast, the current CMS measurements do not exhibit a statistically significant positive correlation at small separations and therefore do not allow a meaningful determination of the model parameters. At present, the CMS data remain compatible with both the proposed decoherence scenario and the absence of an observable initial spin correlation.

The framework developed here demonstrates that the language of open quantum systems provides a natural and quantitative description of the loss of spin correlations during hadronization. The results suggest that the experimentally observed dependence of $P_{\Lambda\bar{\Lambda}}$ on $\Delta R$ can be understood in terms of environmental decoherence acting on an initially correlated quark pair. In the future, it would be meaningful to further extend this approach to account for the system's evolution in color space, as recently proposed in~\cite{Liu:2026ees}.

\noindent\emph{\textbf{Acknowledgments:}} J.B. is grateful to R. Venugopalan and W. Gong for early discussions on the study of hyperon spin correlations. This work is supported by the European Research Council under project ERC-2018-ADG-835105 YoctoLHC; by Maria de Maeztu excellence unit grant CEX2023-001318-M and project PID2023-152762NB-I00 funded by MICIU/AEI/10.13039/501100011033; and by ERDF/EU. It has received funding from Xunta de Galicia (CIGUS Network of Research Centres). B.W. acknowledges the support of the Ram\'{o}n y Cajal program with the Grant No. RYC2021-032271-I and the support of Xunta de Galicia under the ED431F 2023/10 project. This work has been partially funded by the Eric \& Wendy Schmidt Fund for Strategic Innovation through the CERN Next Generation Triggers project under grant agreement number SIF-2023-004. I. C.
acknowledges the support of the Axudas de apoio á etapa predoutoral program (Ref. ED481A-2024-075).

\appendix
\section{Microscopic derivation of open system master evolution equations} \label{app:MicroscopicDerivation}

Consider a quantum system $S$ coupled to an environment $B$. The total Hamiltonian of $S+B$ can be written as
\begin{align}
H = H_S + H_B + H_I \, ,
\end{align}
where $H_S$ and $H_B$ denote the free Hamiltonians of the system and bath, respectively, while $H_I$ describes their interaction. The state of the combined system is represented by the density operator $\rho_{\mathrm{tot}}(t)$, which evolves according to the von Neumann equation in the Schrödinger picture
\begin{align}
\frac{d\rho_{\mathrm{tot}}(t)}{dt}=-i\left[H,\rho_{\mathrm{tot}}(t)\right]\, .
\end{align}
Although the evolution of the total system is unitary, the subsystem of interest is described by the reduced density matrix
\begin{align}
\rho_{\mathrm{tot}}(t)=\mathrm{Tr}_B\left[\rho_{\mathrm{tot}}(t)\right],
\end{align}
obtained by tracing over the environmental degrees of freedom. In general, the dynamics of $\rho_{\mathrm{tot}}(t)$ are non-unitary and depend on the entire previous history of the system. Consequently, the exact reduced evolution is typically non-Markovian and difficult to solve for.

To derive a tractable equation of motion, it is convenient to work in the interaction picture. Defining the free Hamiltonian $H_0 = H_S + H_B$, the interaction picture interaction Hamiltonian is
\begin{align}
H_I(t)=e^{iH_0 t}H_Ie^{-iH_0 t}\, .
\end{align}
The von Neumann equation then becomes
\begin{align}\label{eq:fff}
\frac{d\rho_{\mathrm{tot}}(t)}{dt}=
-i\left[H_I(t),\rho_{\mathrm{tot}}(t)\right],
\end{align}
where now $\rho(t)$ is in the interaction picture as well.
Integrating Eq.~\eqref{eq:fff} and taking the partial trace over the bath degrees of freedom leads to an exact integro-differential equation for the reduced density matrix: 
\begin{align}
\frac{d\rho(t)}{dt}=
-\int_0^t ds\,\mathrm{Tr}_B\left\{
    \left[
         H_I(t),\left[H_I(s),\rho_{\mathrm{tot}}(s)\right]
    \right]
\right\}\, ,
\end{align}
where we have assumed
\begin{align}
\mathrm{Tr}_B(\left[
    H_I(t),\rho_{\mathrm{tot}}(0)
\right])=0\, .
\end{align}
So far, no approximations have been taken, and the evolution remains non-local in time and depends on the full system-bath state. Using the Born approximation
\begin{align}
\rho_{\mathrm{tot}}(t)\simeq\rho(t)\otimes\rho_B,
\end{align} 
where $\rho_B$ is a stationary bath state satisfying
\begin{align} \label{eq:StationaryState}
[H_B,\rho_B]=0 \, ,
\end{align}
the evolution of the system density matrix reduces to
\begin{align}
\frac{d\rho(t)}{dt}=
-\int_0^t ds\,\mathrm{Tr}_B\left\{
    \left[
        H_I(t),\left[H_I(s),\rho(s)\otimes\rho_B
    \right]
    \right]
\right\} \, .
\end{align}

The Markov approximation allows the integrand $\rho(s)$ to be first replaced by $\rho(t)$. This approximation is valid when the correlation functions decay on a characteristic timescale $\tau_B$ that is much shorter than the characteristic timescale $\tau_S$ over which the reduced density matrix evolves, as the environment rapidly loses memory of previous interactions with the system. With this approximation, one obtains the so-called Redfield master equation~\cite{Redfield1957}. Nonetheless, the evolution still depends on an explicit choice for the initial preparation at time $t=0$. We therefore make the substitution $s\rightarrow t-s$, which results in an exactly Markovian system:
\begin{align}\label{eq:ttt}
\frac{d\rho(t)}{dt}=
-\int_0^\infty ds\,\mathrm{Tr}_B
\left\{
    \left[
    H_I(t),
        \left[
            H_I(t-s),\rho(t)\otimes\rho_B
        \right]
    \right]
\right\}.
\end{align}
Although Eq.~\eqref{eq:ttt} is Markovian, it does not necessarily preserve complete positivity of the density matrix. This can be ensured via a further approximation, writing the interaction Hamiltonian as
\begin{align}
H_I=
\sum_\alpha A_\alpha\otimes B_\alpha\, , 
\end{align}
where $A_\alpha$ and $B_\alpha$ are system and bath operators, respectively. We decompose the system operators into eigenoperators of the free system Hamiltonian. Introducing the spectral decomposition
\begin{align}
A_\alpha
=
\sum_\omega
A_\alpha(\omega)\, ,
\end{align}
the operators $A_\alpha(\omega)$ satisfy
\begin{align}
[H_S,A_\alpha(\omega)]
=
-\omega
A_\alpha(\omega)\, ,
\end{align}
 it follows that
\begin{align}
H_I(t)=
\sum_{\alpha,\omega} e^{-i\omega t} A_\alpha(\omega)\otimes B_\alpha(t).
\end{align}
Let us denote the reservoir average of the relevant correlations of the bath operators as
\begin{align}
\braket{B_\alpha(t)B_\beta(t-s)}=
\mathrm{Tr}_B(B_\alpha(t)B_\beta(t-s)\rho_B).
\end{align}
If Eq.~\eqref{eq:StationaryState} is satisfied, then $\braket{B_\alpha(t)B_\beta(t-s)}=\braket{B_\alpha(s)B_\beta(0)}$, showing that these quantities do not depend on time.
 As the interaction-picture evolution contains oscillatory terms with frequency $\omega-\omega'$, when the frequencies mismatch, such terms average out over timescales in which $\rho_S$ varies appreciably. Thus, we further neglect such oscillatory contributions (secular approximation). Under the combined assumptions of weak coupling, rapidly decaying bath correlations, and the secular approximation, the reduced dynamics thus assume the Lindblad form shown in Eq.~\eqref{eq:Lindblad}.

\section{Replacement quantum channel}
As an alternative to the depolarizing description developed in this work, we also investigated a more direct representation of hadronization in terms of quantum channels that map \textbf{any} initial $s\bar{s}$ spin state onto the experimentally inferred $\Lambda\bar{\Lambda}$ density matrix through a replacement channel.

Such a channel provides an exact description of the observed state transformation and represents an extreme limit of information loss, in which the details of the initial spin configuration are entirely erased by the hadronization process, leaving only the final reconstructed two-particle state.
\subsection{Overview of replacement channel}
The replacement map sends every input state $\rho$ to a fixed output density matrix, independently of the initial state, i.e., given a target final state $\rho_f$, the action of a replacement channel is~\cite{Watrous_2018}
\begin{align}
\mathcal{E}_{\rho_f}(\rho)=\rho_f \, .
\end{align}
Such a map is completely positive and trace preserving, and can be written explicitly in Kraus form by diagonalizing the target state,
\begin{align} \label{eq:FinalRhoDecomposition}
\rho_f=\sum_i \lambda_i \ket{\phi_i}\bra{\phi_i},
\qquad
\sum_i \lambda_i =1 ,
\end{align}
where $\lambda_i$ and $\ket{\phi_i}$ are the eigenvalues and eigenvectors of $\rho_f$. Introducing an arbitrary orthonormal basis ${\ket{\varphi_j}}$ for the input Hilbert space, one possible Kraus representation is
\begin{align} \label{eq:ReplacementKrausOperators}
K_{ij}=\sqrt{\lambda_i}\ket{\phi_i}\bra{\varphi_j}\, ,
\end{align}
where the Kraus index $a$ has been replaced by the composite index $(i,j)$. Replacement channels, therefore, describe an extreme form of information loss: all dependence on the input density matrix is erased, and the system is reset to a prescribed final state. 

This construction is also useful because it allows the channel to be characterized through its Choi--Jamiolkowski state~\cite{Choi:1975nug}, which allows one to determine whether the channel is entanglement breaking, entanglement annihilating, or capable of preserving non-classical correlations. For a channel acting on a Hilbert space $\mathcal{H}$ of dimension $d$,
let $\{\ket{n}\}_{n=1}^{d}$ denote an arbitrary orthonormal basis of $\mathcal{H}$. Introducing the maximally entangled state
\begin{align}
\ket{\Omega}
=
\frac{1}{\sqrt{d}}
\sum_{n=1}^{d}
\ket{n}\otimes\ket{n},
\end{align}
with $\langle\Omega|\Omega\rangle=1$, the Choi--Jamiolkowski state of the channel is defined as
\begin{align}
\Upsilon_{\mathcal{E}}
&=
d\left(\mathbb{1}\otimes \mathcal{E}\right)
\left(\ket{\Omega}\bra{\Omega}\right)
\notag\\
&=
\sum_{n,n'=1}^{d}
\left(\ket{n}\bra{n'}\right)
\otimes
\mathcal E\!\left(\ket{n}\bra{n'}\right).
\end{align}
Equivalently, if the channel is given in Kraus form, its Choi matrix can be written as~\cite{Homa:2025}
\begin{align} \label{eq:ChoiMatrix}
   \Upsilon_{\mathcal{E}} = \sum_{a} { \kketbbra{ K_{a} }{ K_{a} } }, 
\end{align}
Here $\kket{\,}$ denotes a vectorized matrix. In the case of a bipartite quantum channel with $\{K_a\}\in\mathcal{H}_\mathcal{A}\otimes\mathcal{H}_\mathcal{B}$, the Choi matrix acts on the Hilbert space  $(\mathcal{H}_\mathcal{A}\otimes\mathcal{H}_\mathcal{B})\otimes(\mathcal{H}_{\mathcal{A}'}\otimes\mathcal{H}_{\mathcal{B}'})$. 
The Choi matrix is hermitian as it is a sum of rank-1 projectors. It is also unique. Any rank-1 decomposition of $\Upsilon$ in the form in Eq.~(\ref{eq:ChoiMatrix}) gives us a valid set of Kraus operators.

The separability properties of $\Upsilon_{\mathcal{E}}$ provide a diagnostic of the channel itself. Indeed, it has been proven~\cite{PhysRevLett.86.544} that if and only if the Choi matrix is separable across the $\mathcal{A}\mathcal{A}'|\mathcal{B}\mathcal{B}'$ cut, then its corresponding quantum channel is also separable. A necessary condition for the separability of the Choi matrix across a certain cut is that it must satisfy the PPT criterion~\cite{Peres:1996}, which states that, if $\Upsilon$ is separable across, say, the $\mathcal{A}\mathcal{A}'|\mathcal{B}\mathcal{B}'$ cut, then all the eigenvalues of $\Upsilon^{T_{\mathcal{A}\mathcal{A}'}}$ are non-negative, where $T_{\mathcal{A}\mathcal{A}'}$ denotes the partial transpose, obtained by applying transposition on $\mathcal{H}_\mathcal{A}\otimes\mathcal{H}_\mathcal{A}'$ while applying the identity on $\mathcal{H}_\mathcal{B}\otimes\mathcal{H}_\mathcal{B}'$.

On the other hand, a channel is entanglement-breaking~\cite{Horodecki_2003} if its Choi state is separable with respect to the reference--system bipartition. For a replacement channel, one has $\mathcal{E}_{\rho_f}(\ket{n}\bra{n'})=\delta_{nn'}\rho_f$ and, accordingly, finds
\begin{align}
\Upsilon_{\mathcal{E}_{\rho_f}}
=
\mathbb{1}\otimes \rho_f \, .
\end{align}
Hence, the channel is entanglement-breaking: any entanglement between the input system and an external reference is erased. This statement should be distinguished from the question of whether the output two-particle state $\rho_f$ itself is entangled. For a replacement channel, the map is entanglement annihilating (that is, it destroys any entanglementent \textit{within} the system) precisely when the target state $\rho_f$ is separable. Since the present application involves two qubits, this latter question can be tested through the positivity of the partial transpose of $\rho_f$, just like it was explained for $\Upsilon$, the difference being that, whereas for the Choi matrix the PPT criterion provides only a necessary condition for separability, for $\rho_f$ it provides both a necessary and sufficient condition.

\subsection{Replacement channels from the measured final state}

In order to build the channel $\mathcal{E}_{\rho_{\Lambda\bar\Lambda}}(\rho)$, we first decompose our final density matrix as in Eq.~(\ref{eq:FinalRhoDecomposition}). The eigenstates of $\rho_{\Lambda\bar\Lambda}$ are the Bell states
\begin{equation}
\begin{aligned}
   \ket{ B_{00}} = \ket{\Phi^{+}} &= \frac{1}{\sqrt{2}} ( \ket{00} + \ket{11} )\, , \\
   \ket{ B_{01}} = \ket{\Psi^{+}} &= \frac{1}{\sqrt{2}} ( \ket{01} + \ket{10} )\, , \\
   \ket{ B_{10}} = \ket{\Phi^{-}} &= \frac{1}{\sqrt{2}} ( \ket{00} - \ket{11} )\, , \\
   \ket{ B_{11}} = \ket{\Psi^{-}} &= \frac{1}{\sqrt{2}} ( \ket{01} - \ket{10} ),
\end{aligned}
\end{equation}
and its eigenvalues
\begin{equation}
\begin{aligned}
    \lambda_{\Phi^{\pm}} = \frac{1}{4} ( 1 \pm C_x \mp C_y + C_z ), \\
    \lambda_{\Psi^{\pm}} = \frac{1}{4} ( 1 \pm C_x \pm C_y - C_z ).
\end{aligned}
\end{equation}
Following Eq.~(\ref{eq:ReplacementKrausOperators}), we use these results to construct the following sixteen Kraus operators:
\begin{align}
    K_{ij,kl} = \sqrt{ \lambda_{B_{ij}} } \ketbra{B_{ij}}{B_{kl}} \, ,
    \label{eq:BellKrausOp}
\end{align}
which act on any initial density matrix as
\begin{align}
    \mathcal{E}_{\rho_{\Lambda\bar\Lambda}}(\rho) = \sum_{i,j,k,l} K_{ij,kl} \rho K_{ij,kl}^\dagger = \rho_{\Lambda\bar\Lambda}.
\end{align}

\begin{figure}[ht!]
    \centering
\includegraphics[width=0.9\linewidth]{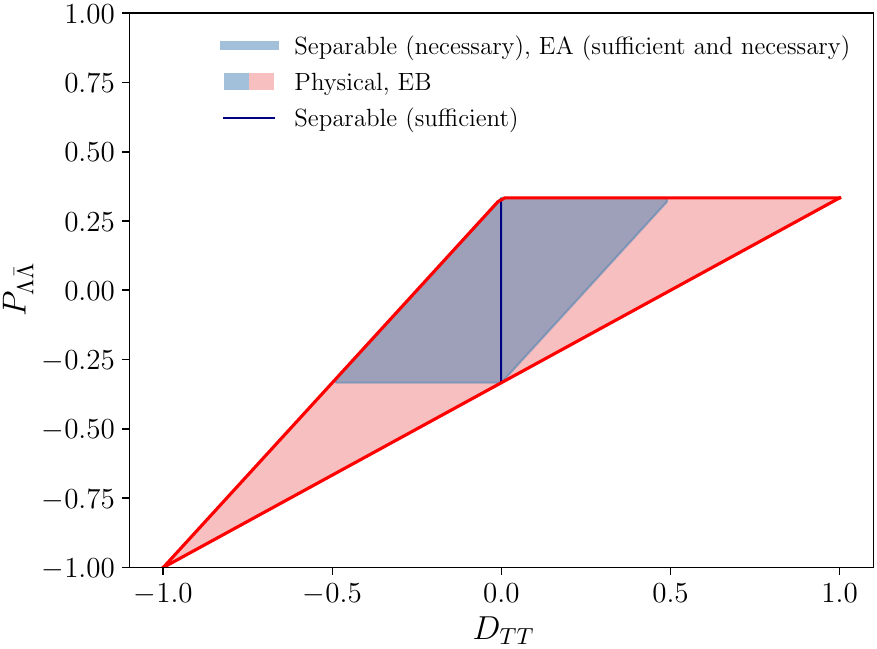}
    \caption{Classification of the replacement, exhibiting the same structure as the state entanglement characterization considered in~\cite{vonKuk:2025kbv}.}
    \label{fig:ChannelClassification}
\end{figure}

Now we proceed to classify our quantum channel following the discussion in the previous section. We have already seen that the replacement channel is always entanglement-breaking. Next, we check the separability of the channel $\mathcal{E}_{\rho_f}$. The eigenvalues of the partial transpose of $\Upsilon_{\mathcal{E}_{\rho_f}}$ across the $\mathcal{A}\mathcal{A}'|\mathcal{B}\mathcal{B}'$ cut are
\begin{align}
    \kappa_{B_{ij}} = \frac{1}{2}\left( -\lambda_{B_{ij}} + \sum_{ij\neq kl}{\lambda_{B_{kl}}}\right).
\end{align}
For the PPT criterion to be satisfied, it must be the case that $\kappa_{B_{ij}}\geq 0$. Therefore, the requirement that $\mathcal{E}_{\rho_f}$ be separable imposes a restriction on the $\lambda_{B_{ij}}$ and, consequently, on the coefficients $C_i$. Using Eq.~(\ref{eq:PLambda}) we find
\begin{align}
    \kappa_{\Phi^{\pm}} &= \frac{1}{4}( 1 - 3P_{\Lambda\bar\Lambda} + 2D_{TT} )\, , \nn
    \kappa_{\Psi^{+}} &= \frac{1}{4}( 1 + 3P_{\Lambda\bar\Lambda} - 4D_{TT}  ) \, ,\nn
    \kappa_{\Psi^{-}} &= \frac{1}{4}( 1 + 3P_{\Lambda\bar\Lambda} ),
\end{align}
from where we can derive the following conditions
\begin{align} \label{eq:PPT}
    \frac{ 4D_{TT} - 1 }{ 3 } \le &P_{\Lambda\bar\Lambda} \le \frac{ 2D_{TT} + 1 }{ 3 }\, , \\
    &P_{\Lambda\bar\Lambda} \ge -\frac{1}{3}\, .
\end{align}
Moreover, positivity of $\rho_{\Lambda\bar\Lambda}$ implies 
\begin{align} \label{eq:DConditions}
   2|D_{TT}| \leq 1 - D_{LL}\, , \nn 
   -1\leq D_{LL} \leq 1\, ,
\end{align}
which restricts the possible values of $P_{\Lambda\bar\Lambda}$ to the range
\begin{align}
    -1 \leq P_{\Lambda\bar\Lambda} \leq\frac{1}{3}.
\end{align}
Combining these conditions, we find that for the channel to be separable, it must be the case that 
\begin{align}
    -\frac{1}{3} \leq\ P_{\Lambda\bar\Lambda} \leq\frac{1}{3}.
\end{align}
Fig.~\ref{fig:ChannelClassification} shows the region where these inequalities are satisfied.

To find a sufficient condition for separability, we first compute the reshuffled Choi matrix $\Upsilon^R \in (\mathcal{H}_{\mathcal{A}}\otimes\mathcal{H}_{\mathcal{A}'})\otimes(\mathcal{H}_{\mathcal{B}}\otimes\mathcal{H}_{\mathcal{B}'})$, which has the form
\begin{align}
    \Upsilon^R = \begin{pmatrix}
        \Delta_0 & \Gamma & 0 & 0 \\
        \Gamma^T & \Delta_1 & 0 & 0 \\
        0 & 0 & \Delta_0 & \Gamma \\
        0 & 0 & \Gamma^T & \Delta_1
    \end{pmatrix},
\end{align}
where $\Delta_0$ and $\Delta_1$ are diagonal matrices and 
\begin{widetext}
  \begin{align}
    \Gamma = \frac{1}{2} \begin{pmatrix}
        0 & \lambda_{\Phi^{+}}-\lambda_{\Phi^{-}} & 0 & 0 \\
        \lambda_{\Psi^{+}}-\lambda_{\Psi^{-}} & 0 & 0 & 0 \\
        0 & 0 & 0& \lambda_{\Phi^{+}}-\lambda_{\Phi^{-}} \\
        0 & 0 & \lambda_{\Psi^{+}}-\lambda_{\Psi^{-}} & 0 
    \end{pmatrix}\, ,
\end{align}  
\end{widetext}
with $\lambda_{\Phi^{+}}-\lambda_{\Phi^{-}}=0$ always. A sufficient condition for the channel to be separable is that $\Gamma=0$, which happens when $
    \lambda_{\Psi^{+}} - \lambda_{\Psi^{-}} = 0 \Rightarrow D_{TT} = 0$,
with $D_{LL}$ taking any value between $-1$ and $1$. Indeed, in this case, the reshuffled Choi matrix can be written as
\begin{align}
    \Upsilon^R_{\mathcal{E}_{\rho_f}} &=
    (\mathbb{1}_2\otimes\ketbra{0}{0})\otimes\Delta_0 + (\mathbb{1}_2\otimes\ketbra{1}{1})\otimes\Delta_1
    =
    \notag\\
    &=
    \sum_{l=0}^1 \sum_{\alpha,\beta} { 
        \kketbbra{ K_{l,\alpha}^{\mathcal{A}} }{ K_{l,\alpha}^{\mathcal{A}} } \otimes \kketbbra{ K_{l,\beta}^{\mathcal{B}} }{ K_{l,\beta}^{\mathcal{B}} }
    },
\end{align}
which means that the original Choi matrix can be expressed as
\begin{align}
    \Upsilon_{\mathcal{E}_{\rho_f}} = \sum_{l=0}^1 \sum_{\alpha,\beta} {
        \kketbbra{K_{l,\alpha}^{\mathcal{A}}\otimes K_{l,\beta}^{\mathcal{B}}}{K_{l,\alpha}^{\mathcal{A}}\otimes K_{l,\beta}^{\mathcal{B}}}
    }
\end{align}
and its corresponding quantum channel is, therefore,
\begin{align}
    \mathcal{E}_{\rho_{\Lambda\bar\Lambda}}(\rho) = 
    \sum_{l=0}^1 \sum_{\alpha,\beta} {
        (K_{l,\alpha}^{\mathcal{A}}\otimes K_{l,\beta}^{\mathcal{B}})\rho(K_{l,\alpha}^{\mathcal{A}}\otimes K_{l,\beta}^{\mathcal{B}}) }\, ,
\end{align}
which is patently separable.
Finally, as it has already been mentioned, we note that the replacement channel is entanglement-annihilating if and only if the final state $\rho_{\Lambda\bar{\Lambda}}$ is separable. On the other hand, for a replacement channel, separability of the final state is a necessary condition for the separability of the channel, since a separable channel can not map separable states to entangled ones. We then conclude that the region where the channel is entanglement-annihilating coincides with the region where the channel meets the necessary condition to be separable (see Fig.~\ref{fig:ChannelClassification}).

\bibliographystyle{JHEP-2modlong.bst}

\bibliography{Lib.bib}

\clearpage
\onecolumngrid  
\appendix

\end{document}